\documentclass[letterpaper,nofootinbib,prd,amsmath,twocolumn]{revtex4-1}

\usepackage{epsfig}
\usepackage{hyperref}

\begin{document}

\title{Action Principle for the Generalized Harmonic Formulation of General Relativity}
\author{J.~David Brown}
\affiliation{Department of Physics, North Carolina State University,
Raleigh, NC 27695 USA}

\begin{abstract}
An action principle for the generalized harmonic formulation of general relativity is presented. The 
action is a functional of the spacetime metric and the gauge source vector. An action principle 
for the Z4 formulation of general relativity has been proposed recently by 
Bona, Bona--Casas and Palenzuela (BBP). The relationship between the generalized harmonic action and 
the BBP action is discussed in detail. 
\end{abstract}
\maketitle

\section{Introduction}
Einstein's equations can be expressed as an initial value problem using the familiar 3+1 
splitting \cite{Smarr:York,ADM:Witten}.
For numerical applications, one must supplement the 3+1 equations with coordinate conditions. Typically the full set of 
partial differential equations (PDE's) obtained in this way is not well posed (see, for example, 
Refs.~\cite{Reula:1998ty,Calabrese:2002ej}). 
Equations that are not well posed can be used for formal analyses, but they cannot be used for 
numerical applications. Generalized harmonic (GH) gravity is a reformulation of Einstein's theory as a set of 
PDE's that is well posed. The GH equations are currently in use by a number of numerical relativity 
groups (see, for example, Refs.~\cite{Pretorius:2006tp,Lindblom:2005qh,Anderson:2007kz}).

Einstein completed development of his general theory of relativity in a series of papers published in 1915 
\cite{Einstein:1915by,Einstein:1915bz,Einstein:1915ca}.
In the same year, Hilbert derived the field equations for general relativity by postulating a simple action principle 
motivated by  general covariance \cite{Hilbert:1915tx}. 
The Hilbert action provides an economical and efficient way to 
define the theory. Throughout history, physicists have used variational principles as a way of organizing and simplifying their 
descriptions of dynamical systems. Most physicists view the action as fundamental, and the classical equations of motion 
as derived quantities. The action  is typically the starting point for a quantum analysis. 

Generalized harmonic gravity is a generalization of general relativity in the harmonic gauge. 
The generalization to (in principle) arbitrary gauge conditions was first pointed out by Friedrich 
\cite{Friedrich:GH}, and later by Garfinkle \cite{Garfinkle:2001ni}. To my knowledge, the action for 
GH gravity has not been previously discussed. An action for general relativity in harmonic gauge was written 
down by Stone and Kuch{\v a}r \cite{Stone:1992nz}. Their action was not complete in the sense that the 
harmonic coordinate conditions were not included among the equations of motion. Other efforts to write 
well--posed formulations of Einstein's equations in terms of a variational principle can be found 
in Refs.~\cite{Brown:2008cca,Hilditch:2010wp,Bona:2010is}. 

Although generalized harmonic gravity is not a new theory, merely a reformulation of general relativity, the 
action principle presented in this paper provides 
a new perspective on the generalized harmonic system. This new perspective can help us understand the connection 
between GH gravity and other formulations of the Einstein equations. 
The GH action can serve as the basis for practical numerical calculations using variational or symplectic
integrators \cite{Brown:2005jc,Frauendiener:2008bt,Richter:2009ff}. 

It is worth noting that any system of equations can be derived from a variational principle: Simply multiply each equation 
by an undetermined multiplier, add them together, and integrate over spacetime (for PDE's) or time 
(for ordinary differential equations). 
Such an action principle does not add any insights, and probably has no practical benefit. 
What we want in an action principle is an encoding of the equations of motion without the addition of 
extra unphysical variables that do not appear in the original differential equations. 
Not all systems of equations can be derived from such a variational principle. For example, it appears that the 
Baumgarte--Shapiro--Shibata--Nakamura (BSSN) formulation 
of Einstein's theory \cite{Shibata:1995we,Baumgarte:1998te} cannot be derived from an action principle 
using only the BSSN variables. 

The action for GH gravity is presented in Sec.~\ref{section2}. 
One of the features that emerges from this analysis is the need to introduce a background connection. The GH equations 
are not usually written in terms of a background connection; equivalently, the background connection is usually 
set to zero. In numerical relativity applications this can be justified by choosing the background connection to be flat and 
interpreting the coordinates as Cartesian. Note that Kreiss, Reula, Sarbach and Winicour introduce a background 
metric in their studies of constraint--preserving boundary conditions for the generalized harmonic equations 
\cite{Kreiss:2008ig,Winicour:2009dr}. 

The Z4 system is a reformulation of Einstein's equations that, with suitable coordinate conditions, is 
well--posed \cite{Bona:2003fj,Bona:2003qn}. Bona, Bona--Casas and Palenzuela (BBP) have recently proposed an 
action principle for Z4 \cite{Bona:2010is}. In 
Sec.~\ref{section3} I discuss the relationship between the equations of motion obtained from the BBP action 
and the Z4 equations, and point out their differences. The differences are sublte and interesting. The key difference stems from the fact that the BBP 
action, like the familiar Palatini variational principle \cite{Misner:1974qy}, 
treats the spacetime metric and the connection as independent variables. As a result, the Ricci tensor that 
appears in the equations of motion for the BBP action is constructed from the independent connection and not from the 
Christoffel symbols. It is not clear whether or not the equations of motion for the BBP action  have the nice properties 
of the Z4 equations. This shortcoming 
of the BBP variational principle can be corrected if we make a suitable change of variables and 
eliminate the connection as an independent variable. The result is the GH action. 

In the Appendix I discuss the inverse problem of the calculus of variations. This provides a 
complementary perspective to the conclusions reached in Sec.~\ref{section3}. In particular I argue that the 
equations of motion that follow from the BBP functional are not equivalent to the Z4 equations. A brief summary  
is contained in Sec.~\ref{section5}. 

\section{Action for GH gravity}\label{section2}
Let $g_{\mu\nu}$ denote the spacetime metric and $\Gamma^\alpha{}_{\mu\nu}$ denote the 
metric--compatible connection (the Christoffel symbols). Let $\tilde\Gamma^\alpha{}_{\mu\nu}$ 
denote a background connection that is torsion--free and therefore symmetric in its lower indices. 
We will use the shorthand notation 
\begin{eqnarray}\label{DelGammaDef}
	\Delta\Gamma^\alpha{}_{\mu\nu} & \equiv & \Gamma^\alpha{}_{\mu\nu} - \tilde\Gamma^\alpha{}_{\mu\nu}
	\nonumber\\
	& = & \frac{1}{2} g^{\alpha\beta} \left( \tilde\nabla_\mu g_{\nu\beta} + \tilde\nabla_\nu g_{\mu\beta} 
		- \tilde\nabla_\beta g_{\mu\nu} \right) 
\end{eqnarray}
for the difference between these connections. 
The symbol $\tilde\nabla_\mu$ denotes the covariant 
derivative built from $\tilde\Gamma^\sigma{}_{\mu\nu}$. Note that $\Delta\Gamma^\alpha{}_{\mu\nu}$ 
is a type $1\choose 2$ tensor. Throughout this paper indices are raised and lowered with the metric $g_{\mu\nu}$. Thus, 
for example, $\Delta\Gamma_{\mu\beta}{}^\beta = g_{\mu\nu} g^{\alpha\beta} \Delta\Gamma^\nu{}_{\alpha\beta}$. 

The generalized harmonic constraints are defined by
\begin{equation}\label{CDef}
  {\cal C}_\mu \equiv H_\mu + \Delta\Gamma_{\mu\beta}{}^\beta  \ ,
\end{equation} 
where $H_\mu$ is the gauge source vector. 
The  action for generalized harmonic gravity is the following functional of $g_{\mu\nu}$ and 
$H_\mu$:\footnote{The background connection $\tilde\Gamma^\sigma{}_{\mu\nu}$ appears in the action as an external 
field and is not varied.}
\begin{equation}\label{GHaction}
  S[g_{\mu\nu},H_\mu] =  
		\int d^4x \, \sqrt{-g} g^{\mu\nu} \left[ R_{\mu\nu}
		- \frac{1}{2} {\cal C}_\mu {\cal C}_\nu \right]  \ .
\end{equation}
Here, $R_{\mu\nu}$ is the Ricci tensor built from $\Gamma^\alpha{}_{\mu\nu}$. Also,  
units have been chosen so that $16\pi G = 1$, where $G$ is Newton's constant. 

Before continuing, let me comment on the presence of the background connection. 
Since the Lagrangian must be a scalar density, then ${\cal C}_\mu$ must be a covector. 
If we omit $\tilde\Gamma^\sigma{}_{\mu\nu}$ from the definition (\ref{CDef}), then $H_\mu$ 
must transform in such a way that $H_\mu + g_{\mu\nu} g^{\alpha\beta} \Gamma^\nu{}_{\alpha\beta}$ 
is a covector. Recall that under a change of spacetime coordinates, the transformation 
rule for the Christoffel symbols $\Gamma^\nu{}_{\alpha\beta}$ includes an inhomogeneous term. This 
inhomogeneous term, which is multiplied by $g_{\mu\nu} g^{\alpha\beta}$, must be canceled by a 
corresponding term from $H_\mu$. It follows that the transformation rule for 
$H_\mu$ must include an inhomogeneous term that depends on the metric. It is not possible for the 
transformation of $H_\mu$ to depend on the metric unless $H_\mu$ itself depends on the metric. However, 
for the moment, we would like to treat the metric $g_{\mu\nu}$ and the gauge source $H_\mu$ as independent 
variables in the action principle. For this reason, the background connection is needed to 
compensate for the inhomogeneity in the transformation rule for $\Gamma^\alpha{}_{\mu\nu}$. 

With the background connection included in the definition of the constraints ${\cal C}_\mu$, the 
gauge source $H_\mu$ is a covector. Although it is not logically {\em necessary} for $H_\mu$ to transform 
as a covector, as long as we are willing to give it a suitable dependence on $g_{\mu\nu}$, it is at least 
{\em convenient} for $H_\mu$ to transform as a covector.
For example, we might find that a certain source $H_\mu$ works 
well for numerical simulations of black holes with a code that uses a Cartesian coordinate grid. 
Perhaps we would like to reproduce these results with a code that uses a spherical coordinate grid. If $H_\mu$ 
is a covector, we can easily determine the correct form for the gauge source in spherical coordinates. 

Also observe that for most practical numerical applications, it would be natural to choose $\tilde \Gamma^\sigma{}_{\mu\nu}$ 
to be the flat connection. In this case the background connection components $\tilde \Gamma^\sigma{}_{\mu\nu}$ 
would be zero in Cartesian coordinates, but nonzero in spherical coordinates. 

Now consider the variation of the action (\ref{GHaction}). The functional derivatives of $S[g_{\mu\nu},H_\mu]$ are 
\begin{widetext}
\begin{subequations}\label{deltaGHaction}
\begin{eqnarray}
	\frac{\delta S}{\delta H_\mu} & = & -\sqrt{-g} \, {\cal C}^\mu \ ,\\
	\frac{\delta S}{\delta g_{\mu\nu}} & = & 
	-\sqrt{-g} \left[ G^{\mu\nu} - \nabla^{(\mu} {\cal C}^{\nu)} + {\cal C}^{(\mu}\Delta\Gamma^{\nu)\beta}{}_\beta
	- {\cal C}^\sigma\Delta\Gamma_\sigma{}^{\mu\nu} - \frac{1}{2} {\cal C}^\mu {\cal C}^\nu 
	+ \frac{1}{2} g^{\mu\nu} \nabla_\sigma {\cal C}^\sigma 
	+ \frac{1}{4} g^{\mu\nu} {\cal C}_\sigma {\cal C}^\sigma \right]  \ ,
\end{eqnarray}
\end{subequations}
\end{widetext}
where $G^{\mu\nu} \equiv R^{\mu\nu} - Rg^{\mu\nu}/2$ is the Einstein tensor. Parentheses around indices 
denote symmetrization. Note that $\nabla_\mu$ is the covariant 
derivative built from the Christoffel symbols $\Gamma^\alpha{}_{\mu\nu}$. It is related to the background 
covariant derivative by $\nabla_\mu V_\nu = \tilde\nabla_\mu V_\nu - \Delta\Gamma^\sigma{}_{\mu\nu} V_\sigma$, which holds for any covector $V_\mu$. 
The vacuum Einstein equations are obtained by setting the functional derivatives (\ref{deltaGHaction}) to zero. 
Equation~(\ref{deltaGHaction}a) tells us that ${\cal C}^\mu = 0$; hence ${\cal C}^\mu$ are constraints
for the generalized harmonic system. With ${\cal C}^\mu = 0$, 
Eq.~(\ref{deltaGHaction}b) reduces to the vacuum Einstein equations $G^{\mu\nu} = 0$. 
Matter fields can be included in a straightforward way.

A convenient form of the equations of motion is obtained by choosing 
$\sqrt{-g}g^{\mu\nu}$ and $-\sqrt{-g} H^\mu$ as independent variables in the variational 
principle, rather than $g_{\mu\nu}$ and $H_\mu$. This leads to the vacuum equations 
\begin{subequations}\label{GHeqns}
\begin{eqnarray} 
	0 & = & \frac{\delta S}{\delta (-\sqrt{-g}H^\mu)} = {\cal C}_\mu \ , \\
	0 & = & \frac{\delta S}{\delta (\sqrt{-g}g^{\mu\nu})} = 
	R_{\mu\nu} - \tilde\nabla_{(\mu}{\cal C}_{\nu)} 
	+ \frac{1}{2} {\cal C}_\mu {\cal C}_\nu  \ .
\end{eqnarray}
\end{subequations} 
Note that the generalized harmonic equation are usually written in the form $R_{\mu\nu} - \nabla_{(\mu}{\cal C}_{\nu)} = 0$. Neither 
Eq.~(\ref{deltaGHaction}b) nor Eq.~(\ref{GHeqns}b) is identical to the usual equation. The differences are terms proportional 
to the constraints ${\cal C}_\mu$. These terms depend on the choice of independent variables and are not particularly important. 
As we will see, the presence or absence of
these terms does not affect the properties that makes the generalized harmonic equations useful. 

The equations of motion (\ref{deltaGHaction}) are equivalent to 
Einstein's equations. Of course, this assumes that each equation holds for all time. In particular, 
the constraints ${\cal C}_\mu = 0$ must hold for all time. We would like to re--interpret these equations 
as an initial value problem. For this purpose we follow the analysis of Lindblom, Scheel, Kidder, Owen  and 
Rinne \cite{Lindblom:2005qh}, 
and derive two key results from
Eq.~(\ref{deltaGHaction}b). Let $n_\mu$ denote the unit normal to a foliation of spacetime by spacelike hypersurfaces, 
and let $h_{\mu\nu} = g_{\mu\nu} + n_\mu n_\nu$ denote the metric induced on these hypersurfaces. 
The first result is obtained by contracting Eq.~(\ref{deltaGHaction}b) with $n_\nu$, which  yields
\begin{widetext}
\begin{equation}\label{Gmununnu}
	 G^{\mu\nu}n_\nu - \frac{1}{2} n^\sigma\nabla_\sigma {\cal C}^\mu 
	= \frac{1}{2} ( h^{\mu\sigma} n_\rho - h^\sigma_\rho n^\mu)\nabla_\sigma {\cal C}^\rho
	-  n_\nu \left[ {\cal C}^{(\mu}\Delta\Gamma^{\nu)\beta}{}_\beta - {\cal C}^\sigma \Delta\Gamma_\sigma{}^{\mu\nu} 
	- \frac{1}{2} {\cal C}^\mu {\cal C}^\nu + \frac{1}{4} g^{\mu\nu} {\cal C}_\sigma {\cal C}^\sigma \right] \ .
\end{equation}
The second result is obtained by letting the covariant derivative $\nabla_\nu$ act on 
Eq.~(\ref{deltaGHaction}b) and using the Ricci identity. This gives
\begin{equation}\label{nablanuGmunu}
	\nabla^\sigma \nabla_\sigma {\cal C}^\mu = -R^\mu_\sigma {\cal C}^\sigma 
	+ 2\nabla_\nu \left[ {\cal C}^{(\mu}\Delta\Gamma^{\nu)\beta}{}_\beta - {\cal C}^\sigma \Delta\Gamma_\sigma{}^{\mu\nu} 
	- \frac{1}{2} {\cal C}^\mu {\cal C}^\nu + \frac{1}{4} g^{\mu\nu} {\cal C}_\sigma {\cal C}^\sigma \right] \ ,
\end{equation}
\end{widetext}
where the term $\nabla_\nu G^{\mu\nu}$ has been set to zero by the contracted Bianchi identity. 

The first term on the left--hand side of Eq.~(\ref{Gmununnu})  
is the Hamiltonian and momentum constraints, which we denote 
${\cal M}^\mu \equiv G^{\mu\nu} n_\nu $. The second term on the left--hand side is proportional to 
$n^\sigma\nabla_\sigma {\cal C}^\mu = (\partial_t {\cal C}^\mu - \beta^i \partial_i {\cal C}^\mu)/\alpha 
+ n^\sigma \Gamma^\mu{}_{\sigma\nu} {\cal C}^\nu$. Each of the terms on the right--hand side of Eq.~(\ref{Gmununnu})  
is proportional to the constraints ${\cal C}^\mu$ or their spatial derivatives. 
It follows that Eq.~(\ref{Gmununnu}) has the form
\begin{equation}\label{Gmununnu2}
	{\cal M}^\mu - \frac{1}{2\alpha} \partial_t {\cal C}^\mu =  
	\bigl\{ {\hbox{terms $\sim {\cal C}$, $\partial_i {\cal C}$}} \bigr\} \ ,
\end{equation}
where $\partial_i C$ denotes spatial derivatives of ${\cal C}^\mu$. 

Now consider the initial value problem. Equation (\ref{Gmununnu2}) tells us that if ${\cal C}^\mu$ and ${\cal M}^\mu$ vanish 
initially, then $\partial_t {\cal C}^\mu$ vanishes initially. Then Eq.~(\ref{nablanuGmunu}) implies that ${\cal C}^\mu$ 
will remain zero throughout the evolution defined by Eq.~(\ref{deltaGHaction}b). In turn, Eq.~(\ref{Gmununnu2}) 
tells us that ${\cal M}^\mu$  will remain zero throughout the evolution. The same conclusion can be reached 
by splitting the derivatives in Eq.~(\ref{nablanuGmunu}) 
into space and time. Together with Eq.~(\ref{Gmununnu2}) one finds the results
\begin{subequations}\label{MandCeqns}
\begin{eqnarray}
	\partial_t {\cal C}^\mu & = & \bigl\{ {\hbox{terms $\sim {\cal M}$, ${\cal C}$, $\partial_i {\cal C}$}}\bigr\} \ ,\\
	\partial_t {\cal M}^\mu & = & \bigl\{ {\hbox{terms $\sim {\cal M}$, $\partial_i {\cal M}$, 
	${\cal C}$, $\partial_i {\cal C}$, $\partial_i\partial_j {\cal C}$}}\bigr\} \ .
\end{eqnarray}
\end{subequations}
These equations are consequences of Eq.~(\ref{deltaGHaction}b) alone. 
Therefore,  if the constraints ${\cal C}^\mu$ and ${\cal M}^\mu$ vanish initially, then the evolution equation (\ref{deltaGHaction}b) 
will maintain the values ${\cal C}^\mu = {\cal M}^\mu = 0$ throughout the evolution. 

Observe that Eqs.~(\ref{deltaGHaction}b) and (\ref{GHeqns}b) are not equivalent. 
If we take the trace reversed version of Eq.~(\ref{GHeqns}b) and raise its indices, 
the result differs from Eq.~(\ref{deltaGHaction}b) by terms that are linear and quadratic 
in the constraints ${\cal C}_\mu$. The difference does not depend 
on derivatives of the ${\cal C}$'s. As a result, the arguments that led to Eqs.~(\ref{MandCeqns}) hold for 
the evolution equation (\ref{GHeqns}b) as well. In fact, we are free to drop any terms in 
Eqs.~(\ref{deltaGHaction}b) or (\ref{GHeqns}b) that are linear or quadratic 
in the constraints. 

The discussion above shows that the relations (\ref{MandCeqns}) hold for any equation of the form 
\begin{equation}\label{newGHeom}
	R_{\mu\nu} - \tilde\nabla_{(\mu} {\cal C}_{\nu)} = \bigl\{ {\hbox{terms $\sim {\cal C}$}} \bigr\}  \ .
\end{equation}
The terms proportional to ${\cal C}_\mu$ can include, for example, constraint damping terms. 
From the definition of the Ricci tensor we have
\begin{widetext}
\begin{eqnarray}\label{RicciDef}
	R_{\mu\nu} & = & -\frac{1}{2} g^{\alpha\beta} \tilde\nabla_\alpha \tilde\nabla_\beta 
		g_{\mu\nu} + \tilde\nabla_{(\mu} \Delta\Gamma_{\nu)\beta}{}^\beta - g^{\alpha\beta} 
		\tilde R^\sigma{}_{\alpha\beta(\mu} g_{\nu)\sigma} \nonumber\\
	& & + g^{\alpha\beta} \left[ -\Delta\Gamma_{\sigma\alpha\beta}\Delta\Gamma^\sigma{}_{\mu\nu} 
		+ 2\Delta\Gamma^\sigma{}_{\alpha(\mu}\Delta\Gamma_{\nu)\beta\sigma} 
		+ \Delta\Gamma^\sigma{}_{\mu\alpha} \Delta\Gamma_{\sigma\nu\beta} \right] \ ,
\end{eqnarray}
where $\tilde R^\sigma{}_{\alpha\beta\mu}$ is the Riemann tensor built from 
the background connection $\tilde \Gamma^\alpha{}_{\mu\nu}$. 
Then the evolution equation (\ref{newGHeom}) becomes 
\begin{eqnarray}\label{GHevolutioneqn}
	g^{\alpha\beta} \tilde\nabla_\alpha \tilde\nabla_\beta g_{\mu\nu} 
	& = & - 2\tilde\nabla_{(\mu} H_{\nu)}  - 2g^{\alpha\beta} 
		\tilde R^\sigma{}_{\alpha\beta(\mu} g_{\nu)\sigma} \nonumber\\
	& &  + 2g^{\alpha\beta} \left[ -\Delta\Gamma_{\sigma\alpha\beta}\Delta\Gamma^\sigma{}_{\mu\nu} 
		+ 2\Delta\Gamma^\sigma{}_{\alpha(\mu}\Delta\Gamma_{\nu)\beta\sigma} 
		+ \Delta\Gamma^\sigma{}_{\mu\alpha} \Delta\Gamma_{\sigma\nu\beta} \right] 
		+ \bigl\{ {\hbox{terms $\sim {\cal C}$}} \bigr\}  \ .
\end{eqnarray}
\end{widetext}
This is a wave equation for each component of the spacetime metric. The initial value problem for the GH system is
described as follows: Specify initial data for $g_{\mu\nu}$ and $H_\mu$ that satisfies 
${\cal C}^\mu  = {\cal M}^\mu = 0$, then evolve the metric 
with the wave equation (\ref{GHevolutioneqn}). Observe that the gauge source vector $H_\mu$ 
is freely specifiable, apart from the restriction ${\cal C}_\mu = 0$ at the initial time.  

\section{BBP Action}\label{section3}
The functional 
\begin{equation}\label{Z4action}
  S[g_{\mu\nu},Z_\mu,\bar\Gamma^\sigma{}_{\mu\nu}] = \int d^4x \, \sqrt{-g} g^{\mu\nu} \left[ \bar R_{\mu\nu}
   + 2\bar\nabla_\mu Z_\nu \right] 
\end{equation}
was proposed by Bona, Bona--Casas and Palenzuela (BBP) in Ref.~\cite{Bona:2010is} as an 
action principle for the Z4 formulation of general relativity.
This action is a functional of the spacetime metric $g_{\mu\nu}$, a covariant vector $Z_\mu$, and a 
torsion--free  connection $\bar\Gamma^\sigma{}_{\mu\nu}$. The covariant derivative 
$\bar\nabla_\mu$ is built from this connection. Likewise the Ricci tensor that appears in the Lagrangian 
is defined by 
\begin{equation}\label{Rbardef}
	\bar R_{\mu\nu} = \partial_\sigma \bar\Gamma^\sigma{}_{\mu\nu} - \partial_\nu \bar\Gamma^\sigma{}_{\mu\sigma} 
	+ \bar\Gamma^\rho{}_{\mu\nu} \bar\Gamma^\sigma{}_{\rho\sigma} 
	- \bar\Gamma^\rho{}_{\mu\sigma} \bar\Gamma^\sigma{}_{\nu\rho} \ .
\end{equation}
(This definition differs slightly from that of Ref.~\cite{Bona:2010is}. As defined here, $\bar R_{\mu\nu}$ is not 
necessarily symmetric.) We will frequently use the abbreviation 
\begin{equation}\label{Omegadef}
    \Omega^\sigma{}_{\mu\nu}  \equiv  \bar\Gamma^\sigma{}_{\mu\nu} - \Gamma^\sigma{}_{\mu\nu}   \ ,
\end{equation}
for the difference between the connection $\bar\Gamma^\sigma{}_{\mu\nu}$ and the Christoffel symbols 
$\Gamma^\sigma{}_{\mu\nu}$. 
Note that indices are raised and lowered with $g_{\mu\nu}$ and 
its inverse. Thus, for example, $\bar\Gamma_{\sigma\mu\nu} \equiv g_{\sigma\rho}\bar\Gamma^\rho{}_{\mu\nu}$. 

Variation of the BBP action yields the vacuum equations
\begin{subequations}\label{Z4equations}
  \begin{eqnarray}
   0 & = & \frac{\delta S}{\delta(\sqrt{-g} g^{\mu\nu})} = \bar R_{(\mu\nu)} + 2\bar\nabla_{(\mu} Z_{\nu)} \ ,\\
   0 & = & \frac{1}{\sqrt{-g}}\frac{\delta S}{\delta Z_\mu} = -2 \Omega^{\mu\sigma}{}_\sigma \ ,\\
   0 & = & \frac{1}{\sqrt{-g}}\frac{\delta S}{\delta \bar\Gamma^\sigma{}_{\mu\nu}} = \Omega^\rho{}_{\rho\sigma} \, g^{\mu\nu} 
       - 2\Omega^{(\mu\nu)}{}_\sigma \nonumber\\
       & & \qquad\qquad\qquad + \delta^{(\mu}_\sigma \Omega^{\nu)\rho}{}_\rho - 2 Z_\sigma g^{\mu\nu} \ .
\end{eqnarray}
\end{subequations}
For convenience, we have chosen the independent variables to be $\sqrt{-g}g^{\mu\nu}$, $Z_\mu$, and 
$\bar\Gamma^\sigma{}_{\mu\nu}$. It is straightforward to include matter fields. 

Equations (\ref{Z4equations}), which follow from the BBP action, 
are  equivalent to Einstein's general relativity. To show this, we first solve 
Eq.~(\ref{Z4equations}c) for $\bar\Gamma^\sigma{}_{\mu\nu}$. 
This equation can be rearranged to give
\begin{equation}\label{dSdbarGamma}
	2g^{\mu\nu} Z_\sigma = \Omega^\rho{}_{\rho\sigma} g^{\mu\nu} - 2\Omega^{(\mu\nu)}{}_\sigma 
	+ \delta^{(\mu}_\sigma \Omega^{\nu)\rho}{}_\rho  \ .
\end{equation}
By setting $\nu = \sigma$ we obtain  
\begin{equation}\label{dSdbarGamma1}
	\Omega^{\mu\rho}{}_\rho = \frac{4}{3} Z^\mu \ ,
\end{equation}
where the spacetime dimension is assumed to be 4. 
Now take the trace over the indices $\mu$ and $\nu$ in Eq.~(\ref{dSdbarGamma}) to yield 
\begin{equation}\label{dSdbarGamma2}
	\Omega^\rho{}_{\rho\mu} = \frac{10}{3} Z_\mu  \ .
\end{equation}
Putting the results (\ref{dSdbarGamma}--\ref{dSdbarGamma2}) together gives 
\begin{equation}\label{dSdbarGamma3}
	\Omega_{\mu\nu\sigma} + \Omega_{\nu\mu\sigma}  = \frac{4}{3} \left( Z_\sigma g_{\mu\nu} 
		+ Z_{(\mu} g_{\nu)\sigma} \right) \ .
\end{equation}
Now write down two more copies of this equation with index replacements $\mu\to\nu$, $\nu\to\sigma$, $\sigma\to\mu$ 
in the first copy and $\mu\to\sigma$, $\nu\to\mu$, $\sigma\to\nu$ in the second. Add the second copy to Eq.~(\ref{dSdbarGamma3}), 
then subtract the first copy.  This yields 
\begin{equation}\label{dSdbarGammasol}
	\Omega^\sigma{}_{\mu\nu} = \frac{4}{3} \delta^\sigma_{(\mu} Z_{\nu)} 
\end{equation} 
for the solution of Eq.~(\ref{Z4equations}c). 

The vacuum equation of motion (\ref{Z4equations}b) implies
\begin{equation}
	\Omega^{\mu\rho}{}_\rho = 0 \ .
\end{equation}
With the result (\ref{dSdbarGammasol}) we see that Eqs.~(\ref{Z4equations}b) and (\ref{Z4equations}c), together, 
have the solution 
\begin{subequations}\label{twoeqnssoln}
\begin{eqnarray}
	Z_\mu & = & 0 \ , \\
	\Omega^\sigma{}_{\mu\nu} & = & 0 \ .
\end{eqnarray}
\end{subequations}
The second of these equations tells us that the connection $\bar\Gamma^\sigma{}_{\mu\nu}$ is equal to the 
Christoffel symbols. The results (\ref{twoeqnssoln}) show that the  equation of motion
(\ref{Z4equations}a) is equivalent to the vacuum Einstein equations, $R_{\mu\nu} = 0$. 

The Z4 equations are usually written as $R_{\mu\nu} + 2\nabla_{(\mu}Z_{\nu)} = 0$ and $Z_\mu = 0$. 
The equation $R_{\mu\nu} + 2\nabla_{(\mu}Z_{\nu)} = 0$  has the same key
properties as Eq.~(\ref{deltaGHaction}b) or (\ref{GHeqns}b) for the GH system. By the same analysis that led to 
Eqs.~(\ref{MandCeqns}), one can show that the equation $R_{\mu\nu} + 2\nabla_{(\mu}Z_{\nu)} = 0$  implies
\begin{subequations}\label{MandZeqns}
\begin{eqnarray}
	\partial_t Z^\mu & = & \bigl\{ {\hbox{terms $\sim {\cal M}$, $Z$, $\partial_i Z$}}\bigr\}  \ , \\
	\partial_t {\cal M}^\mu & = & \bigl\{ {\hbox{terms $\sim {\cal M}$, $\partial_i {\cal M}$, 
	$Z$, $\partial_i Z$, $\partial_i\partial_j Z$}}\bigr\}  \ .
\end{eqnarray}
\end{subequations}
Thus, if $Z_\mu = 0$ and ${\cal M}_\mu = 0$ initially, then $Z_\mu$ and ${\cal M}_\mu$ will remain zero throughout 
the evolution defined by $R_{\mu\nu} + 2\nabla_{(\mu}Z_{\nu)} = 0$.

Unfortunately, the equation (\ref{Z4equations}a) that comes from the BBP action does not appear to have this 
property, for two reasons. First, the trace--reversed Ricci tensor 
$\bar G_{\mu\nu} \equiv \bar R_{\mu\nu} - g_{\mu\nu}\bar R_{\alpha\beta} g^{\alpha\beta}/2$, built with the connection 
$\bar\Gamma^\sigma{}_{\mu\nu}$, does not satisfy the contracted Bianchi identities. Second, the
Hamiltonian and momentum constraints are not equivalent to the normal projections of $\bar G_{\mu\nu}$. 
The argument showing that $Z_\mu$ and ${\cal M}_\mu$ will 
remain zero, assuming they are zero initially, does not obviously hold for the equation
$\bar R_{(\mu\nu)} + 2\bar\nabla_{(\mu}Z_{\nu)} = 0$. 

Since Eqs.~(\ref{Z4equations}b) and (\ref{Z4equations}c), together, imply 
$\bar\Gamma^\sigma{}_{\mu\nu} = \Gamma^\sigma{}_{\mu\nu}$ and $Z_\mu = 0$, we are
free to replace the connection $\bar\Gamma^\sigma{}_{\mu\nu}$ with the Christoffel symbols 
$\Gamma^\sigma{}_{\mu\nu}$ when solving the equations of motion. It follows that 
Eqs.~(\ref{Z4equations}) are equivalent to the system
\begin{subequations}\label{Z4eqsreduced}
  \begin{eqnarray}
   0 & = &  R_{\mu\nu} + 2\nabla_{(\mu} Z_{\nu)} \ ,\\
   0 & = & - 2 Z_\sigma g^{\mu\nu} \ ,
\end{eqnarray}
\end{subequations}
obtained by setting $\bar\Gamma^\sigma{}_{\mu\nu} = \Gamma^\sigma{}_{\mu\nu}$ in 
Eqs.~(\ref{Z4equations}a) and (\ref{Z4equations}c). These are the Z4 equations. However, these equations do not 
appear to coincide with the extrema of any action functional. In other words, there is no 
functional (to my knowledge) whose functional derivatives 
are linear combinations of $R_{\mu\nu} + 2\nabla_{(\mu} Z_{\nu)}$ and $- 2 Z_\sigma g^{\mu\nu}$. 
This point is discussed more thoroughly in the Appendix. 

Note in particular that the functional obtained by setting 
$\bar\Gamma^\sigma{}_{\mu\nu} = \Gamma^\sigma{}_{\mu\nu}$ in the BBP action (\ref{Z4action}) does 
not yield Eqs.~(\ref{Z4eqsreduced}) for its extrema. 
This is an example of a general rule: One cannot always 
reduce an action principle by using results from the equations of motion. Consider an action $S[u,v]$ that is a functional 
of two sets of variables, $u^i$ and $v^a$. If the equations of motion $\delta S/\delta u^i = 0$ 
can be solved for the variables $u^i$ as functions of $v^a$, 
then it is indeed permissible to use the solutions $u^i = u^i(v)$ to eliminate $u^i$ from the action. 
On the other hand, one or more of the 
equations $\delta S/\delta u^i = 0$ might yield, for example, $v^1$ as a function of the other $v$'s and the $u$'s. 
It is {\it not} permissible to use this result to eliminate $v^1$ from the action. 

In light of these remarks, consider the BBP action (\ref{Z4action}) and the equations of motion 
(\ref{Z4equations}). As the result (\ref{dSdbarGammasol}) shows, the equation (\ref{Z4equations}c) has the solution 
\begin{equation}
	\bar\Gamma^\sigma{}_{\mu\nu}  = \Gamma^\sigma{}_{\mu\nu} + \frac{4}{3} \delta_{(\mu}^\sigma Z_{\nu)} \ .
\end{equation}
In this case we have solved the equation $\delta S/\delta\bar\Gamma^\sigma{}_{\mu\nu} = 0$ 
for $\bar\Gamma^\sigma{}_{\mu\nu}$ and we are allowed to use this solution to simplify the action. The result is 
\begin{equation}\label{altaction}
	S[g_{\mu\nu},Z_\mu] = \int d^4x \, \sqrt{-g}g^{\mu\nu} \left[ R_{\mu\nu} - \frac{4}{3} Z_\mu Z_\nu \right] \ ,
\end{equation}
and the equations of motion become 
\begin{subequations}\label{alteqns}
\begin{eqnarray}
	0 & = & \frac{\delta S}{\delta (\sqrt{-g}g^{\mu\nu})} =  R_{\mu\nu} - \frac{4}{3} Z_\mu Z_\nu \ , \\
	0 & = & \frac{1}{\sqrt{-g}} \frac{\delta S}{\delta Z_\mu} = -\frac{8}{3} Z^\mu  \ .
\end{eqnarray}
\end{subequations}
These equations are physically correct---they are equivalent to vacuum general relativity.   
They do not, however, have the form of the usual Z4 equations.

Another option is to solve the equations of motion (\ref{Z4equations}b) and (\ref{Z4equations}c), together, for 
$\bar\Gamma^\sigma{}_{\mu\nu}$ and $Z_\mu$. The solution is listed in Eqs.~(\ref{twoeqnssoln}). If we use these 
results to eliminate $\bar\Gamma^\sigma{}_{\mu\nu}$ and $Z_\mu$ from the action we are left with the Hilbert 
action. The equations of motion are the vacuum Einstein equations which are, of course, 
physically correct. However, they are not the usual Z4 equations. 

The equation of motion (\ref{Z4equations}c), by itself, does not imply 
$\bar\Gamma^\sigma{}_{\mu\nu} = \Gamma^\sigma{}_{\mu\nu}$ due to the presence of the fields $Z_\mu$. We can 
try to eliminate $Z_\mu$ from the functional derivative $\delta S/\delta \bar\Gamma^{\sigma}{}_{\mu\nu}$ by changing the 
independent variables in the action principle. Since a change of independent variables will merely mix 
the equations of motion, 
it will not be possible to eliminate $Z_\mu$ from $\delta S/\delta \bar\Gamma^{\sigma}{}_{\mu\nu}$ unless $Z_\mu$ appears 
undifferentiated in one of the other equations of motion. With a simple modification of the action, the fields $Z_\mu$ 
will appear in the functional derivatives  $\delta S/\delta Z_\mu$. Thus, let
\begin{widetext}
\begin{equation}\label{Z4action2}
  S[g_{\mu\nu},Z_\mu,\bar\Gamma^\sigma{}_{\mu\nu}] = \int d^4x \, \sqrt{-g} g^{\mu\nu} \left[ \bar R_{\mu\nu}
   + 2\bar\nabla_\mu Z_\nu  + \lambda Z_\mu Z_\nu \right] \ ,
\end{equation}
so that the equations of motion become 
\begin{subequations}\label{Z4equations2}
  \begin{eqnarray}
   0 & = & \frac{\delta S}{\delta(\sqrt{-g} g^{\mu\nu})} = \bar R_{(\mu\nu)} + 2\bar\nabla_{(\mu} Z_{\nu)} 
	+ \lambda Z_\mu Z_\nu  \ ,\\
   0 & = & \frac{1}{\sqrt{-g}}\frac{\delta S}{\delta Z_\mu} = -2 \Omega^{\mu\sigma}{}_\sigma 
	+ 2\lambda Z^\mu  \ ,\\
   0 & = & \frac{1}{\sqrt{-g}}\frac{\delta S}{\delta \bar\Gamma^\sigma{}_{\mu\nu}} = \Omega^\rho{}_{\rho\sigma} \, g^{\mu\nu} 
       - 2\Omega^{(\mu\nu)}{}_\sigma + \delta^{(\mu}_\sigma \Omega^{\nu)\rho}{}_\rho - 2 Z_\sigma g^{\mu\nu} \ .
\end{eqnarray}
\end{subequations}
\end{widetext}
Here, $\lambda$ is a constant parameter. 

We can now look for a change of independent variables that will mix the equation of motion (\ref{Z4equations2}b) with
(\ref{Z4equations2}c), and in the process eliminate $Z_\mu$ from the functional derivatives 
$\delta S/\delta \bar\Gamma^{\sigma}{}_{\mu\nu}$.
This is accomplished by replacing $Z_\mu$ with a combination of $\bar\Gamma^\sigma{}_{\mu\nu}$ and a new 
independent variable, a covariant vector that we call $H_\mu$. For example, 
we can replace $Z_\mu$ with the linear combination 
\begin{equation}\label{Zmudef}
	Z_\mu = \frac{1}{\lambda} \bigl( H_\mu + \Omega_\mu{}^\beta{}_\beta \bigr) 
\end{equation}
in the action (\ref{Z4action2}). The resulting equations of motion are 
\begin{subequations}\label{Z4equations3}
  \begin{eqnarray}
   0 & = & \frac{\delta S}{\delta(\sqrt{-g} g^{\mu\nu})} = \bar R_{(\mu\nu)} 
	+ \lambda Z_\mu Z_\nu  + \bigl\{ {\hbox{terms $\sim \Omega^\sigma{}_{\alpha\beta}$}} \bigr\}   \ ,\\
   0 & = & \frac{1}{\sqrt{-g}}\frac{\delta S}{\delta H_\mu} = -\frac{2}{\lambda} \Omega^{\mu\sigma}{}_\sigma 
	+ 2 Z^\mu  \ ,\\
   0 & = & \frac{1}{\sqrt{-g}}\frac{\delta S}{\delta \bar\Gamma^\sigma{}_{\mu\nu}} = \Omega^\rho{}_{\rho\sigma} \, g^{\mu\nu} 
       - 2\Omega^{(\mu\nu)}{}_\sigma  \nonumber\\
       & & \qquad\qquad\qquad{\ }  + \delta^{(\mu}_\sigma \Omega^{\nu)\rho}{}_\rho
	- \frac{2}{\lambda} \Omega_\sigma{}^\rho{}_\rho \, g^{\mu\nu}  \ ,
\end{eqnarray}
\end{subequations}
with $Z_\mu$ given by Eq.~(\ref{Zmudef}). Eq.~(\ref{Z4equations3}c) has the desired property---its solution is 
$\bar\Gamma^\sigma{}_{\mu\nu} = \Gamma^\sigma{}_{\mu\nu}$ (assuming $\lambda \ne 4/3$). 
However, Eq.~(\ref{Z4equations3}a) no longer includes the 
term proportional to $\nabla_{(\mu} Z_{\nu)}$ that characterizes the Z4 equation (\ref{Z4eqsreduced}a). 
This is because the change of variables (\ref{Zmudef}) 
contains derivatives of the metric through the Christoffel symbols. 

We can eliminate the Christoffel symbols $\Gamma^\sigma{}_{\mu\nu}$ 
from the change of variables (\ref{Zmudef}) by replacing them with a 
background connection $\tilde\Gamma^\sigma{}_{\mu\nu}$.  Therefore, let 
\begin{equation}\label{Zmudef2}
	Z_\mu = \frac{1}{\lambda} \bigl( H_\mu + \bar\Gamma_\mu{}^\beta{}_\beta 
	- \tilde\Gamma_\mu{}^\beta{}_\beta \bigr) 
\end{equation}
in the action (\ref{Z4action2}). The equations of motion become 
\begin{widetext}
\begin{subequations}\label{Z4equations4}
  \begin{eqnarray}
   0 & = & \frac{\delta S}{\delta(\sqrt{-g} g^{\mu\nu})} = \bar R_{(\mu\nu)} + 2\bar\nabla_{(\mu}Z_{\nu)} 
	+ \lambda Z_\mu Z_\nu + \bigl\{ {\hbox{terms $\sim (\Omega^{\rho\sigma}{}_{\sigma} - \lambda Z^\rho)$}} \bigr\}  \ ,\\
   0 & = & \frac{1}{\sqrt{-g}}\frac{\delta S}{\delta H_\mu} = -\frac{2}{\lambda} \Omega^{\mu\sigma}{}_\sigma 
	+ 2 Z^\mu  \ ,\\
   0 & = & \frac{1}{\sqrt{-g}}\frac{\delta S}{\delta \bar\Gamma^\sigma{}_{\mu\nu}} = \Omega^\rho{}_{\rho\sigma} \, g^{\mu\nu} 
       - 2\Omega^{(\mu\nu)}{}_\sigma + \delta^{(\mu}_\sigma \Omega^{\nu)\rho}{}_\rho 
	- \frac{2}{\lambda} \Omega_\sigma{}^\rho{}_\rho \, g^{\mu\nu} \ ,
\end{eqnarray}
\end{subequations}
\end{widetext}
where $Z_\mu$ is given by Eq.~(\ref{Zmudef2}). The solution of Eq.~(\ref{Z4equations4}c) 
is $\Omega^\sigma{}_{\mu\nu} = 0$ for $\lambda \ne 4/3$, 
and we are allowed to use $\bar\Gamma^\sigma{}_{\mu\nu} = \Gamma^\sigma{}_{\mu\nu}$ in the action to eliminate 
$\bar\Gamma^\sigma{}_{\mu\nu}$. In the process, the definition (\ref{Zmudef2}) becomes 
$Z_\mu = {\cal C}_\mu/\lambda$, where ${\cal C}_\mu$ is the generalized harmonic constraint (\ref{CDef}). 
The action becomes
\begin{equation}\label{Z4asGHaction}
	S[g_{\mu\nu},H_\mu] = \int d^4x \,\sqrt{-g} g^{\mu\nu} \left[ R_{\mu\nu} + \frac{1}{\lambda}
	{\cal C}_\mu {\cal C}_\nu  \right] \ ,
\end{equation}
where the term proportional to $\nabla_\mu Z_\nu$ has been integrated to the boundary and discarded.  
The equations of motion are 
\begin{subequations}\label{Z4asGHequations}
  \begin{eqnarray}
   0 & = & \frac{\delta S}{\delta(\sqrt{-g} g^{\mu\nu})} =  R_{\mu\nu} 
      + \frac{2}{\lambda} \nabla_{(\mu} {\cal C}_{\nu)}
	+ \frac{1}{\lambda} {\cal C}_\mu {\cal C}_\nu \nonumber\\ & & \qquad\qquad\qquad{\ } +  
      \bigl\{ {\hbox{terms $\sim {\cal C}^\sigma$}} \bigr\} \ ,\\
   0 & = & \frac{1}{\sqrt{-g}}\frac{\delta S}{\delta H_\mu} =  \frac{2}{\lambda} {\cal C}^\mu  \ .
\end{eqnarray}
\end{subequations}
When $\lambda = -2$ these are the GH equations and Eq.~(\ref{Z4asGHaction}) is the GH action. 

The preceding analysis shows that we are naturally led to the GH action when we attempt to reformulate 
the BBP action without the connection $\bar\Gamma^\sigma{}_{\mu\nu}$. The GH action (\ref{GHaction}) can 
be obtained directly from the BBP action (\ref{Z4action}) by the change of variables 
\begin{equation}
	Z_\mu = -\frac{1}{2} {\cal C}_\mu + \frac{1}{8} \Omega_\mu{}^\rho{}_\rho \ .
\end{equation}
With this definition, the BBP action becomes 
\begin{widetext}
\begin{equation}\label{anotherZ4toGH}
	S[g_{\mu\nu},H_\mu,\bar\Gamma^\sigma{}_{\mu\nu}] 
	= \int d^4x \, \sqrt{-g} g^{\mu\nu} \left[ \bar R_{\mu\nu} - \bar\nabla_\mu {\cal C}_\nu 
	+ \frac{1}{4} \bar\nabla_\mu \Omega_\nu{}^\rho{}_\rho \right] \ .
\end{equation}
\end{widetext}
The equation of motion $\delta S/\delta\bar\Gamma^\sigma{}_{\mu\nu} = 0$ has the solution 
\begin{equation}
	\bar\Gamma^\sigma{}_{\mu\nu} = \Gamma^\sigma{}_{\mu\nu} - \delta_{(\mu}^\sigma {\cal C}_{\nu)} \ .
\end{equation}
Substituting this result into the action (\ref{anotherZ4toGH}) and discarding a boundary term yields 
the GH action (\ref{GHaction}).

\section{Summary}\label{section5}
The action for the generalized harmonic formulation of general relativity has the remarkably simple form 
displayed in Eq.~(\ref{GHaction}). This action can be used as the starting point for further 
developments, such as the Hamiltonian formulation of GH gravity. We can also use the action to develop  
variational and symplectic integration schemes. The BBP action presented in 
Ref.~\cite{Bona:2010is} is closely related to the GH action, but the equations of motion that follow from 
the BBP action are not obviously equivalent to the Z4 equations. 
After a change of variables, the independent connection $\bar\Gamma^\sigma{}_{\mu\nu}$ can be 
eliminated from the BBP action, reducing it to the GH action.  

\begin{acknowledgments}
This work was supported by 
NSF Grant PHY--0758116. I would like to thank Lee Lindblom for insightful comments at the early 
stages of this work, and Meng Cao for valuable technical help. I would also like to thank 
Carlos Palenzuela for helpful comments. 
\end{acknowledgments}

\appendix*
\section{The inverse problem of the calculus of variations}
The problem of finding an action for the GH (or Z4) equations is an example of the 
inverse problem of the calculus of variations. 
This subject has a long history \cite{AndersonThompson}. In its most basic form, the inverse problem 
of the calculus of variations can be stated as follows. Given a set of differential 
equations $E^A(\phi,\partial\phi,\ldots) = 0$ for the variables $\phi^A$, does there exist a functional 
$S[\phi]$ whose functional derivatives are $E^A(\phi,\partial\phi,\ldots)$? If so, is the 
functional unique? The index $A$ runs from $1$ to $N$ and  $\partial\phi$ represents 
the partial derivatives of the dependent variables $\phi^A$ with respect to the independent variables. 
For ordinary differential equations, there is only one independent variable; for 
partial differential equations, there are two or more independent variables. 
The dots in  $E^A(\phi,\partial\phi,\ldots)$ represent higher order derivatives of $\phi^A$. 

An acceptable action functional for the GH or Z4 equations does not need to reproduce the differential equations 
identically. It is  
acceptable if the functional derivatives of the action are a linear combination of $E^A$. 
This formulation of the inverse problem of the calculus of variations is often referred to as the 
variational multiplier problem \cite{AndersonThompson,Henneaux}. Thus, given a 
system $E^A(\phi,\partial\phi,\ldots) = 0$, we seek a functional $S[\phi]$ that satisfies 
\begin{equation}\label{varmultprob}
     M^{AB}(\phi,\partial\phi,\ldots) \,\frac{\delta S[\phi]}{\delta\phi^B} = E^A(\phi,\partial\phi,\ldots)
\end{equation}
where $M^{AB}$ is an invertible matrix  that depends on $\phi^A$ and its derivatives. 
Equation (\ref{varmultprob}) says that the expressions $E^A$ are linear combinations 
of the functional derivatives of $S[\phi]$.

The inverse problem of the calculus of variations assumes that the action is a functional only of 
those variables $\phi^A$ that appear in the system of equations $E^A = 0$. (It also assumes that 
the number of equations is equal to the number of variables.) 
As an alternative, consider the functional $S[\phi,\Lambda] = \int \Lambda_A E^A(\phi,\partial\phi,\ldots)$ 
of $\phi^A$ and $\Lambda_A$. The functional derivatives of $S[\phi,\Lambda]$ 
include $E^A$. Equivalently, the conditions for the extremization of $S[\phi,\Lambda]$ 
imply $E^A = 0$. In spite of this fact, the functional 
$S[\phi,\Lambda]$ is not considered a valid action for the equations $E^A = 0$ because it 
depends on the extra unphysical variables $\Lambda_A$. 

In the variational multiplier problem (\ref{varmultprob}), $M^{AB}$ can depend on the fields 
$\phi^A$ and their derivatives but it is not allowed to be a differential operator. 
This restriction on $M^{AB}$  is a natural one, since 
we want the functional derivatives of the action to yield the same system of differential equations 
as defined by $E^A = 0$. A derivative operator in $M^{AB}$ can change the differential order of the 
functional derivatives so that the extremum of the action is no longer  equivalent to 
the original differential system. Although this {\it can} happen when $M^{AB}$ contains differential operators, it 
does not always happen. 

Let us consider the consequences of this restriction in the context of the BBP functional (\ref{Z4action}). 
The functional derivatives of the BBP action are displayed in Eqs.~(\ref{Z4equations}). 
A close examination of the
analysis following these equations shows that the functional derivatives 
(\ref{Z4equations}b) and (\ref{Z4equations}c) can be rearranged, by a linear transformation, to 
form the left--hand sides of Eqs.~(\ref{twoeqnssoln}). In other words, there is a matrix $M_1^{AB}$ that 
mixes the functional derivatives of the BBP functional, leading to the result (using matrix notation 
in place of the indices $A$ and $B$) 
\begin{equation}\label{M1mixingA}
      M_1 \left( \frac{\delta S}{\delta \phi} \right) = 
      \left(  \begin{array}{c} \bar R_{(\mu\nu)} + 2\bar\nabla_{(\mu} Z_{\nu)} \\
      Z_\sigma \\
      \Omega^\alpha{}_{\beta\gamma} \end{array} \right)   \ .
\end{equation}
We can use the definitions (\ref{Rbardef}) and (\ref{Omegadef}) to write this result in the form 
\begin{equation}\label{M1mixing}
      M_1 \left( \frac{\delta S}{\delta \phi} \right) 
      = \left( \begin{array}{c} \partial_\rho \bar\Gamma^\rho{}_{\mu\nu} 
      - \partial_{(\mu} \bar\Gamma^\rho{}_{\nu)\rho} + \cdots \\
      Z_\sigma \\
      \Gamma^\alpha{}_{\beta\gamma} - \bar\Gamma^\alpha{}_{\beta\gamma} \end{array} \right)  \ .
\end{equation}
For simplicity, only two terms are displayed in the first row. 

Now we ask whether there exists a further mixing of the functional derivatives that will yield the 
Z4 equations $R_{\mu\nu} + 2\nabla_{(\mu} Z_{\nu)} = 0$ and $Z_\mu= 0$.  The mixture must replace  
derivatives of the background connection $\bar\Gamma^\sigma{}_{\mu\nu}$ with derivatives 
of the Christoffel symbols $\Gamma^\sigma{}_{\mu\nu} $ in the first row of Eq.~(\ref{M1mixing}). 
The matrix that does this is 
\begin{equation}
      M_2 = \left( \begin{array}{ccc} 1 & 0 & \delta_{(\mu}^\beta \delta_{\nu)}^\gamma\partial_\alpha
      - \delta_\alpha^\gamma \delta_{(\mu}^\beta \partial_{\nu)} + \cdots \\
      0 & 1 & 0 \\
      0 & 0 & 1 \end{array} \right)  \ ,
\end{equation}
where each of the $1$'s is an identity tensor. In this example both $M_1$ and $M_2$ are invertible. 
But because $M_2$ contains a derivative operator, the matrix 
$M_2 M_1$ does not qualify as a valid variational multiplier for the inverse problem of the 
calculus of variations. 
The conclusion is that the BBP functional (\ref{Z4action}) does not qualify as an action principle for the Z4 
equations. 

In the present example the differential operator $M_2 M_1$ is invertible, and it does not change the 
differential order of the functional derivatives of the BBP action. 
So perhaps the restriction that $M^{AB}$ should not contain any derivative operators
is too severe. Perhaps the only restriction on $M^{AB}$ should be invertibility. Note, however, that 
if we allow $M^{AB}$ to be a differential operator then there exist action functionals for 
the Z4 equations that are more simple than the BBP functional. For example, 
the action of Eq.~(\ref{altaction}) has functional derivatives
\begin{equation}
      \left( \frac{\delta S}{\delta\phi}\right)  = \left( \begin{array}{c} 
      R_{\mu\nu} - 4 Z_\mu Z_\nu/3 \\
      -8\sqrt{-g} Z^\sigma /3 \end{array} \right) \ ,
\end{equation}
as seen from Eqs.~(\ref{alteqns}). These can be rearranged to give 
\begin{equation}
      M\left( \frac{\delta S}{\delta \phi} \right) = \left( \begin{array}{c} 
      R_{\mu\nu} + 2\nabla_{(\mu} Z_{\nu)} \\
       Z_\rho \end{array} \right) 
\end{equation}
with the invertible matrix 
\begin{equation}
      M = \left( \begin{array}{cc} 1 & -( 2g_{\sigma(\mu}Z_{\nu)} + 3 g_{\sigma(\mu}\nabla_{\nu)} )/(4\sqrt{-g}) \\
      0 & -3g_{\rho\sigma}/(8\sqrt{-g}) \end{array} \right) \ .
\end{equation}
If we allow $M^{AB}$ to mix  $\delta S/\delta\phi^A$ with derivatives of  $\delta S/\delta\phi^A$, 
then by this criterion the functional (\ref{altaction}) would be a valid action principle for Z4. 

The view among researchers who study the inverse problem of the calculus 
of variations is that the variational multiplier should be an invertible matrix that depends only 
on the variables and their derivatives \cite{AndersonThompson}. According to this view, neither the BBP functional 
(\ref{Z4action}) nor the functional of Eq.~(\ref{altaction}) qualify as action principles for Z4. 
The GH functional (\ref{GHaction}), on the other hand, is a valid 
action principle for the GH formulation of general relativity. In particular, the equations 
of motion (\ref{deltaGHaction}) or (\ref{GHeqns}) follow directly from this action and have the 
desired properties discussed in Sec.~\ref{section2}.

\bibliography{references}
\end{document}